\documentclass[reprint]{revtex4-1}

\usepackage{enumerate}
\usepackage{mathrsfs}
\usepackage{dsfont}
\usepackage{graphicx}
\usepackage{hyperref}
\usepackage{amsthm,amsmath}
\usepackage{array}
\newcolumntype{L}{>{$}l<{$}} 

\newcommand{\eps}{\varepsilon}
\newcommand{\RR}{\mathds R}
\newcommand{\EE}{\mathds E}
\newcommand{\NN}{\mathds N}
\newcommand{\ZZ}{\mathds Z}

\newtheorem{proposition}{Proposition}
\theoremstyle{remark}
\newtheorem{remark}{Remark}

\begin{document}

\title[String Method for Generalized Gradient Flows]{String Method for Generalized Gradient Flows: Computation of
  Rare Events in Reversible Stochastic Processes}

\author{Tobias Grafke}
\affiliation{Mathematics Institute, University of Warwick, Coventry
  CV4 7AL, United Kingdom}
\email{T.Grafke@warwick.ac.uk}

\date{\today}

\begin{abstract}
  Rare transitions in stochastic processes can often be rigorously
  described via an underlying large deviation principle. Recent
  breakthroughs in the classification of reversible stochastic
  processes as gradient flows have led to a connection of large
  deviation principles to a generalized gradient structure. Here, we
  show that, as a consequence, metastable transitions in these
  reversible processes can be interpreted as heteroclinic orbits of
  the generalized gradient flow. This in turn suggests a numerical
  algorithm to compute the transition trajectories in configuration
  space efficiently, based on the string method traditionally
  restricted only to gradient diffusions.
\end{abstract}

\maketitle

\section{Introduction}

Meta-stability frequently occurs in nature. A complex stochastic
dynamical system often has multiple locally stable fixed points close
to which it spends the majority of its time. Rarely, and on a much
longer time-scale, fluctuations push the system from one such state to
another. Typical examples include chemical reactions under thermal
noise~\cite{wales:2006}, nucleation~\cite{e-ren-vanden-eijnden:2004},
crystal deformation~\cite{e-ren-vanden-eijnden:2005}, etc. Another
common setup is the coarse-graining of microscopic models in
statistical mechanics, where effective dynamics can be derived
including noise terms as fluctuating
hydrodynamics~\cite{ortiz_de_zarate-sengers:2006}. If the effective
limiting dynamics have multiple fixed points, such as for phase
transitions, or when conditioning on rare observables, the intrinsic
stochastic noise induced by the finiteness of the number of particles
will trigger the rare event, a situation that is quantified by
\emph{macroscopic fluctuation
  theory}~\cite{bertini-de_sole-gabrielli-etal:2015}.

In all these cases, the computation of the most likely transition
pathway is practically achievable if a \emph{large deviation
  principle} (LDP) holds~\cite{freidlin-wentzell:2012}. Whenever
present, the LDP demands that the least unlikely of all transition
scenarios will exponentially dominate all others, reducing the
original stochastic sampling problem to a deterministic optimization
problem. The analytical computation of the corresponding minimizers
(\emph{maximum likelihood pathways}, MLPs) is often impossible, and
their numerical computation leads to a high-dimensional optimization
problem, which for systems with a large number of degrees of freedom
is hard to solve. The computation of the MLP is significantly
simplified for a specific sub-class of stochastic processes: Whenever
the dynamics is a diffusion in a potential landscape with small noise,
the invariant measure of the process is explicitly known from the
potential itself, and MLPs become minimum energy paths. The
computation of transition trajectories is then simplified to the
computation of heteroclinic orbits of the gradient flow, which is
numerically achieved by the string
method~\cite{e-ren-vanden-eijnden:2002, e-ren-vanden-eijnden:2007}.

An evolution driven by a negative gradient of a potential is a
straightforward example of a gradient flow. Recent breakthroughs
allowed phrasing many more reversible systems as (generalized)
gradient flows, starting with recognizing the Wasserstein gradient
structure of the Fokker-Planck equation of It\^o stochastic
differential equations
(SDEs)~\cite{jordan-kinderlehrer-otto:2006}. The relation to large
deviation principles of microscopic particle systems is by now well
understood~\cite{mielke-peletier-renger:2014}. The resulting class of
dynamics is no longer restricted to diffusions or even the Gaussian
case, but applicable e.g.~to jump processes, lattice gas models or
interacting particle systems. The main point of this paper is to show
that under certain conditions, transition trajectories in reversible
stochastic processes are heteroclinic orbits (or their time-reverse)
of the associated generalized gradient flow. This allows us to derive
a generalized string method for the efficient and robust computation
of the MLPs.

\section{Main results}

Let $X^\eps_t \in \mathcal E^\eps$ be a family of continuous time
Markov jump processes (MJPs) in the state spaces $\mathcal E^\eps$. If
$X^\eps_t$
\begin{enumerate}[(i)]
\item fulfills a pathwise LDP in the limit $\eps\to0$ with rate
  function
  \begin{equation*}
    I_T(\phi) = \int_0^T  L(\phi,\dot\phi)\,dt
  \end{equation*}
  for \emph{Lagrangian} $ L$, and corresponding
  \emph{Hamiltonian} $H$ given by the Fenchel-Legendre transform
  \begin{equation*}
    H(\psi,\theta) = \sup_{\eta}\left(\langle\theta, \eta\rangle -  L(\psi,\eta)\right)\,,
  \end{equation*}
  and
\item obeys detailed balance,
\end{enumerate}
then the transition trajectory $\{\phi(\tau)\}$ between two fixed
points $a$ and $b$ in the limit $\eps\to0$ fulfills
\begin{equation*}
  \dot \phi = -\partial_\theta H(\phi,0)
\end{equation*}
between $a$ and the ``relevant saddle'' $z$, and
\begin{equation*}
  \dot \phi = \partial_\theta H(\phi,0)
\end{equation*}
between $z$ and $b$. In particular, the transition trajectory,
i.e.~the trajectory $\phi(t)$ and corresponding $T$ that minimize
$I_T(\phi)$, is described by the heteroclinic orbits (and their
time-reverse) of the \emph{generalized gradient flow} field
$\partial_\theta H(\phi,0)$. This includes the previously known case
of diffusions in a potential $U(\phi)$, where $\dot\phi=\pm \nabla
U(\phi)$.

The main result of the paper is that we can then numerically compute
the transition trajectory with the \emph{generalized string method},
which, starting from an initial guess $\phi_i^0$, where
$i\in\{1,\dots,N\}$ enumerates system copies along the trajectory,
iterates only two simple steps until convergence:
\begin{enumerate}[(i)]
\item Update to temporary states $\tilde\phi_i$,
  \begin{equation*}
    \tilde\phi_i = \phi^{k}_i + \Delta t\, \partial_\theta H(\phi^k_i,0)\,.
  \end{equation*}
\item Reparametrize $\tilde\phi_i$ to obtain next iterates $\phi^{k+1}_i$ that fulfill,
  \begin{equation*}
    \|\phi^{k+1}_{i+1} - \phi^{k+1}_i\| = \mathrm{cst.}\qquad \forall\,i\in\{1,\dots,N\}\,.
  \end{equation*}
\end{enumerate}

The knowledge of the transition trajectory yields not only information
about the most likely transition in the large deviation limit, but
furthermore allows one to estimate the exponential scaling of its
probability. Additionally, it yields the relevant saddle point
(transition state) of the limiting dynamics, i.e.~the points where the
trajectory crosses the separatrix between one basin of attraction and
another, and can be used as a `reaction coordinate' as a basis for
more sophisticated sampling techniques, such as forward flux
sampling~\cite{allen-valeriani-wolde:2009}, cloning or splitting
algorithms~\cite{giardina-kurchan-lecomte-etal:2011,
  cerou-guyader:2007, brehier-gazeau-goudenege-etal:2016} or
importance sampling.

In what follows, we will first derive the main results of the paper in
section~\ref{sec:trans-traj-revers}. Subsequently, in
section~\ref{sec:conn-gener-grad}, the connection to the mathematical
literature on generalized gradient flows is made. In
section~\ref{sec:numer-comp-trans}, the numerical computation of
limiting trajectories is discussed, and the full string method for
generalized gradient flows is introduced. Finally, its capabilities
are demonstrated in section~\ref{sec:examples} on several examples,
including a bi-stable reaction network, a zero range lattice gas model
exhibiting condensation, and the hydrodynamic limit of interacting
particles.

\section{Transition Trajectories in Reversible Markov Jump Processes}
\label{sec:trans-traj-revers}

\subsection{Large deviation principles and transition trajectories}
\label{sec:large-devi-princ}

Let $X^\eps_t \in \mathcal E^\eps$ be a family MJPs in the state
spaces $\mathcal E^\eps$ with generators $\mathcal L^\eps$ and unique
invariant measures $\mu_\infty^\eps$. We say that $X^\eps_t$ fulfills
a pathwise \emph{large deviation principle} (LDP) if, for $\delta>0$
sufficiently small, the probability to observe a sample path close to
a given trajectory $\phi(t)$ fulfills
\begin{equation}
  \mathcal P(\sup_{t\in[0,T]} |X^\eps_t - \phi(t)|<\delta) \asymp \exp(-\eps^{-1} I_T(\psi))\,.
\end{equation}
Here, the sign $\asymp$ stands for asymptotic logarithmic equivalence,
i.e.~that for $\eps\to0$, the logarithm of both sides has the same
limit, and
\begin{equation*}
  I_T(\psi) = \int_0^T L(\psi,\dot\psi)\,dt\,.
\end{equation*}
The quantity $L(\psi,\dot\psi)$ is called the \emph{Lagrangian}, which
admits a corresponding \emph{Hamiltonian} as its Fenchel-Legendre
transform
\begin{equation*}
  H(\psi,\theta) = \sup_\eta (\langle \theta,\eta\rangle - L(\psi,\eta))\,.
\end{equation*}
The probability to observe a transition starting in a neighborhood of
a point $a\in\mathcal E$ to a neighborhood of a point $b\in\mathcal E$
can be obtained by a minimization over suitable trajectories $\psi$,
\begin{equation}
  \label{eq:minimization}
  \mathcal P(a\to b) \asymp \exp(-\eps^{-1} \inf_{\psi\in\mathscr C_a^b} I_T(\psi))
\end{equation}
with $\mathscr C_a^b = \{ \psi(t) \in \mathcal D[0,T] | \psi(0)=a,
\psi(T)=b\}$, $\mathcal E$ denoting the limiting state space, and
$D[0,T]$ the Skorokhod space on $[0,T]$. For a precise definition of
large deviation principles for stochastic processes, see
e.g.~\cite{feng-kurtz:2006}. In the course of this paper, we are
interested in finding explicitly the \emph{minimizer} (or
\emph{instanton}) $\phi$ that solves the minimization
problem~(\ref{eq:minimization}) for given endpoints $a$, $b$,
\begin{equation*}
  I_T(\phi) = \inf_{\psi\in\mathscr C_a^b}
  I_T(\psi)\,,
\end{equation*}
when additionally minimizing over the transition time $T$.  This
minimizer $(\phi(t),T)$ describes the most likely transition
trajectory in the limit $\eps\to0$. It can equivalently be described
as a solution to the Hamilton's equations
\begin{equation}
  \label{eq:hamiltons}
  \dot\phi = \partial_\theta H(\phi,\theta),\qquad \dot \theta =
  -\partial_\phi H(\phi,\theta)\,.
\end{equation}

A special case is the situation of a trajectory $\phi(t)$, such that
$I_T(\phi)=0$. Since $L\ge 0$, these trajectories are necessarily
(global) minimizers, and are called \emph{zero action pathways} or
\emph{relaxation dynamics}. The equivalent dynamics define a
\emph{deterministic} dynamical system as limit of the original
stochastic process which can be interpreted as a law of large numbers
(LLN) or hydrodynamic limit of the MJP. In terms of the Hamiltonian,
they correspond to solutions of (\ref{eq:hamiltons}) with $\theta=0$,
$\dot\theta=0$, and therefore
\begin{equation*}
  \dot\phi = \partial_\theta H(\phi,0)\,.
\end{equation*}
In the situation where the relaxation dynamics have a unique and
stable fixed point $a$, i.e.~$\partial_\theta H(a,0)=0$, and we only
consider trajectories starting from that fixed point, we can define
the \emph{quasipotential} $V(x)$ by
\begin{equation}
  \label{eq:qp}
  V(x) = \inf_{T>0} \inf_{\phi \in \mathcal C_a^x} I_T(\phi)\,.
\end{equation}
The quasipotential loosely quantifies the difficulty of reaching a
point $x$ via the stochastic process and generalizes the notion of a
free energy to non-equilibrium systems. It is connected to the
invariant measure $\mu_\infty^\eps$ of the MJP through the relation
$\lim_{\eps\to0} \eps \log \mu_\infty^\eps(A) = -\inf_{x\in A}
V(x)$~(see~\cite[Chpt 4, Thm 4.3]{freidlin-wentzell:2012} for
details).

\begin{figure}
  \begin{center}
    \includegraphics[width=196pt]{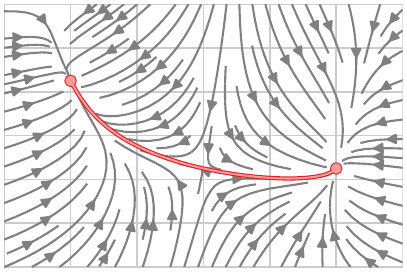}
  \end{center}
  \caption{The minimizing trajectory between two fixed points of a
    gradient flow $\dot\phi=\partial_\theta H(\phi,0)$ is a
    heteroclinic orbit connecting each stable fixed point to the
    relevant saddle point on the separatrix. \label{fig:example}}
\end{figure}

For transition trajectories between two fixed points $a$ and $b$, as
depicted in figure~\ref{fig:example}, we realize that the definition
of the quasipotential~(\ref{eq:qp}) holds true locally in each basin
of attraction. A restriction of the arguments to single basins, and
subsequent `stitching' of $V(x)$ to neighboring
basins~\cite{graham:1987}, proves to be sufficient to describe
bi-stable transitions of the form of figure~\ref{fig:example}, as we
will see in the next section.

\subsection{Adjoint Process and Reversibility}

For a reversible MJP $X_t\in\mathcal E$, the \emph{adjoint process},
generated by the $L^2(\mathcal E, \mu_\infty)$-adjoint of $\mathcal L$
is equal to the process itself, i.e.~for $\rho_\infty$ being the density
of $\mu_\infty$ with respect to counting or Lebesgue measure,
\begin{equation*}
  \mathcal L = \rho_\infty^{-1} \mathcal L^\dagger \rho_\infty\,,
\end{equation*}
where $\mathcal L^\dagger$ is the usual $L^2(\mathcal E)$-adjoint of
$\mathcal L$. Intuitively, the probability of starting at $a$ and
observing the trajectory $\psi(t)$ is equal to the probability of
observing the reverse trajectory $\psi^*(t) = \psi(T-t)$ starting at
$b$. For a family of reversible MJPs $(X_t)^\eps$, with LDT Lagrangian
$L$ and Hamiltonian $H$, this can be expressed as
\begin{equation}
  \label{eq:reverse-prob}
  V(a) + \int_0^T L(\psi,\dot \psi)\,dt = V(b) + \int_0^T L(\psi,-\dot \psi)\,dt\,,
\end{equation}
for any trajectory $\{\psi\}_{t\in[0,T]}$ with $\psi(0)=a,
\psi(T)=b$~(compare~\cite[Thm
  3.3]{mielke-peletier-renger:2014}). 

Written in terms of the Hamiltonian, this translates to
\begin{equation}
  \label{eq:reverse-H}
  H(\psi,\theta) = H(\psi,\nabla V-\theta).
\end{equation}
This is a consequence of~(\ref{eq:reverse-prob}), which implies
\begin{equation}
  \label{eq:L-reverse}
  L(\psi,\dot\psi) = L(\psi,-\dot\psi) + \frac{d}{dt} V(\psi) = L(\psi,-\dot\psi) + \langle\nabla V,\dot \psi\rangle
\end{equation}
and therefore
\begin{align*}
  H(\psi,\theta)&=\sup_{\eta}\left(\langle\eta,\theta\rangle - L(\psi,\eta)\right) \\
  &= \sup_{\eta}\left(\langle\eta,\theta\rangle - \langle\eta,\nabla V\rangle - L(\psi,-\eta)\right) \\
  &= \sup_{-\eta}\left(\langle\eta,\nabla V - \theta\rangle - L(\psi,\eta)\right) \\
  &= H(\psi,\nabla V-\theta)\,.
\end{align*}
Defining an entropy function $S(x)=2V(x)$, this equation can be
rewritten~\cite[Prop~2.1]{mielke-peletier-renger:2014} as
\begin{equation}
  \label{eq:db-H}
  H(\psi, \nabla S - \theta) = H(\psi, \nabla S + \theta)\,.
\end{equation}

\subsection{Minimizing Trajectories of the Action of Reversible Processes}

We now want to focus on the minimizing trajectories of the action for
reversible processes, i.e.~the solutions $(\phi, T)$ to the
minimization problem~(\ref{eq:qp}) under the assumption that the
process is reversible.

Consider first the case that we start at a point $a$ that is a fixed
point of the limiting dynamics, i.e.~$\partial_\theta H(a,0)=0$. We
want to investigate the minimizing transition trajectory to a point
$b$ within the same basin of attraction. Denote by $(\phi,T)$ the
minimizing pair for the quasipotential, equation~(\ref{eq:qp}). Since
$a$ is a fixed point, this necessitates $V(a)=0$ and $T=\infty$. Then,
equation~(\ref{eq:qp}) becomes $V(b) =\int_0^\infty L(\phi,\dot
\phi)\,dt$. By plugging this into equation~(\ref{eq:reverse-prob}),
this shows that $\phi(t)$ is also a trajectory that has
\begin{equation}
  \label{eq:Lstar-is-zero}
  L(\phi,-\dot\phi)=0\,.
\end{equation}
In other words, minimizers originating at the fixed point are
time-reversed relaxation trajectories. More precisely, combining
equations~(\ref{eq:L-reverse}) and~(\ref{eq:Lstar-is-zero}), we have
\begin{equation}
  L(\phi,\dot\phi) = \langle \nabla V(\phi),\dot\phi\rangle
\end{equation}
and therefore $\theta = \frac{\partial L}{\partial\dot\phi} = \nabla V
= 2\nabla S$. For this trajectory, the equation of motion thus reads
\begin{equation}
  \dot\phi = \partial_\theta H(\phi,\nabla V(\phi))\,.
\end{equation}
On the other hand, by differentiating~(\ref{eq:reverse-H}) with
respect to $\theta$, and setting $\theta=0$, we obtain
\begin{equation}
  \partial_\theta H(\phi,\nabla V(\phi)) = -\partial_\theta H(\phi,0)\,.
\end{equation}
We therefore arrive at following the proposition:
\begin{proposition}
  \label{prop:uphill}
  For reversible processes, minimizers $\phi(t)$ of the large
  deviation rate function starting at a fixed point $a$ of the
  limiting dynamics, and ending at any point $b$ within the same basin
  of attraction, fulfill
  \begin{equation}
    \dot\phi = -\partial_\theta H(\phi,0)\,.
  \end{equation}
\end{proposition}

Next, we want to consider the case where both initial and final points
$a$ and $b$ are fixed points with neighboring basins of attraction
that together cover the whole state space. The flow $\partial_\theta
H(\phi,0)$ then defines a separatrix $C$ in a dynamical systems sense
that separates the two basins of $a$ and $b$. This separatrix possibly
contains multiple saddle points $z_i$ with $\partial_\theta
H(z_i,0)=0$. Denote by $z$ the \emph{relevant saddle point}, i.e.~the
point on the separatrix which attains the minimal quasipotential,
$V(z)\le V(x) \forall x\in C$. This point necessarily is also the
point through which the most likely transition between $a$ and $b$
traverses from one basin of attraction to the other: A transition from
$a$ to $b$ must leave the basin of attraction of $a$, which must
happen at $z$ since $z$ minimizes the quasipotential on $C$. The
remaining portion between $z$ and $b$ on the other hand can be
achieved with zero cost: Since $z$ is on the separatrix, there exists
a relaxation trajectory from $z$ to $b$, i.e.~a trajectory $\phi$
connecting $z$ and $b$ such that
\begin{equation}
  \dot\phi = \partial_\theta H(\phi,0)\,.
\end{equation}
Alluding to the case of a diffusion in a potential, this portion of
the trajectory is often called the \emph{downhill} portion, as it
coincides in this case with the direction of maximally decreasing
potential. On the other hand, for the portion between $a$ and $z$,
proposition~\ref{prop:uphill} applies, i.e.~here
\begin{equation}
  \dot\phi = -\partial_\theta H(\phi,0)\,.
\end{equation}
With the same intuition, we call this portion of the trajectory the
\emph{uphill} portion, as the dynamics occur along directions of
maximally increasing potential in the case of diffusion in a
potential. Note though that this intuition breaks down for general
reversible processes. In particular, even though the notion of the
quasipotential replaces the potential, it is not true that relaxation
paths are obeying $\dot\phi = -\nabla V(\phi)$. Instead, relaxation
paths are \emph{generalized gradient flows} in the quasipotential, and
uphill paths are time reversed \emph{generalized gradient flows} in
the quasipotential. Both are not necessarily aligned with the
direction of maximal increase or decrease of the quasipotential.

For the complete transition trajectory, we therefore have, under the
above assumptions ($a$ and $b$ being neighboring fixed points, with
basin of attraction covering the complete state space):
\begin{proposition}
  \label{prop:final}
  For reversible processes, minimizers $\phi(t)$ of the large
  deviation rate function connecting two fixed points $a$ and $b$
  fulfill
  \begin{equation}
    \label{eq:H-only}
    \dot\phi = \partial_\theta H(\phi,0)
  \end{equation}
  in the uphill portion from $a$ to the relevant saddle point $z$, and
  \begin{equation}
    \dot\phi = -\partial_\theta H(\phi,0)
  \end{equation}
  for the downhill portion from $z$ to $b$.
\end{proposition}
Note that both up- and downhill portion of the transition trajectory
are readily available by simple integration, and no reference is made
to the quasipotential. This generalizes the known case of diffusions
in a potential $U(\phi)$, where the minimizers of the rate function
fulfill either
\begin{equation*}
  \dot\phi=-\nabla U(\phi)\qquad\mathrm{or}\qquad\dot\phi = \nabla U(\phi)\,.
\end{equation*}
Intuitively, proposition~\ref{prop:final} states the unsurprising fact
that time-reversal symmetry is obeyed for transition trajectories in
reversible processes. At the same time, it implies that all minimizing
trajectories are \emph{heteroclinic orbits} or time-reversed
heteroclinic orbits of the \emph{generalized gradient flow} induced by
the large deviation principle of $X_t^\eps$. It therefore allows for
an extension of the string method~\cite{e-ren-vanden-eijnden:2002} to
more general situations, which is the main contribution of this
paper. This will be discussed in
section~\ref{sec:numer-comp-trans}.

\section{Connection to generalized gradient flows}
\label{sec:conn-gener-grad}

To make the connection to the mathematical literature, we want to
highlight here the relation of the above considerations to generalized
gradient flows and their connection to large deviation theory. This is
of particular importance in the context of statistical mechanics and
stochastic thermodynamics, where the connection between large
deviations and hydrodynamic
limits~\cite{bertini-de_sole-gabrielli-etal:2015}, statistical
mechanics~\cite{touchette:2009} and non-equilibrium
thermodynamics~\cite{kraaij-lazarescu-maes-etal:2018} is known in
considerable detail.

Consider the space $\mathcal E$ to be a Riemannian manifold, with
tangent bundle $T\mathcal E$ and cotangent bundle $T^\star \mathcal
E$. Define on $\mathcal E$ a convex, continuously differentiable
function $\psi_x(v):T\mathcal E \to \RR^+$ and its Legendre-dual
$\psi^\star_x(w):T^\star\mathcal E \to \RR^+$,
\begin{align*}
  \psi^\star_x(w) &= \sup_{v\in T_x\mathcal E} \left(\langle v,w\rangle - \psi_x(v)\right)\\
  \psi_x(v) &= \sup_{w\in T^\star_x\mathcal E} \left(\langle v,w\rangle - \psi^\star_x(w)\right)\,.
\end{align*}
Additionally we demand $\psi_x(0)=\psi^\star_x(0)=0$. Then,
$(\psi,\psi^\star)$ are called \emph{dissipation potentials} in the
context of thermodynamics.

Furthermore, consider a continuously differentiable function
$S:\mathcal E \to \RR$. Then, any evolution on $\mathcal E$ according
to $\dot x = F(x)$ that fulfills
\begin{equation*}
  \psi_x (F(x)) + \psi^\star_x(-\nabla S(x)) + \langle F(x),\nabla S(x)\rangle = 0
\end{equation*}
is called a \emph{generalized gradient flow} with respect to $(\mathcal E, \psi,
S)$~\cite{mielke-peletier-renger:2014}. This is equivalent to saying
\begin{equation}
  \label{eq:gradient-flow-psi}
  \dot x = \partial_w \psi^\star_x(-\nabla S(x))\,.
\end{equation}
In the case of a gradient diffusion, the dissipation potential
$\psi^\star_x(w)$ is quadratic in $w$ (the quadratic dependence being
a consequence of the Gaussianity of the noise, the metric implied by
its covariance). In that case, $\partial_w\psi^\star_x(w)$ is linear
in its argument, and thus the flow~(\ref{eq:gradient-flow-psi}) is a
traditional gradient flow proportional to $\nabla S(x)$. Allowing for
generic dissipation potentials $\psi^\star_x(w)$ explains why the
flows~(\ref{eq:gradient-flow-psi}) are called \emph{generalized}
gradient flows.

The connection to large deviations is made when taking
\begin{equation}
  \label{eq:psi-choice}
  \psi^\star_x(\theta) = H(x,\theta+\nabla S(x)) - H(x,\nabla S(x))\,,
\end{equation}
for a large deviation Hamiltonian $H(\psi,\theta)$ of a reversible
process. The choice~(\ref{eq:psi-choice}) fulfills the assumptions of
$(\psi,\psi^\star)$ to be dissipation potentials, and a gradient flow
can be constructed out of the large deviation principle, or,
equivalently, the optimization problem of large deviations can be
interpreted as the variational formulation of a generalized gradient
flow.

Finally, equation~(\ref{eq:gradient-flow-psi}) with the
choice~(\ref{eq:psi-choice}), leads to
\begin{equation*}
  \dot x = \partial_\theta H(x,0)
\end{equation*}
as gradient flow. As we will see in the examples in
section~\ref{sec:examples}, non-Gaussian stochastic processes lead to
Hamiltonians that are not quadratic in their conjugate momentum, and
therefore to generalized instead of traditional gradient flows.

\section{Numerical Computation of the Transition Trajectory}
\label{sec:numer-comp-trans}

With the realization of proposition~\ref{prop:final}, implementing a
string method for generalized gradient flows in the spirit of the
original string method~\cite{e-ren-vanden-eijnden:2002,
  e-ren-vanden-eijnden:2007} becomes straightforward.

Consider a reversible MJP $X_t^\eps\in\mathcal E^\eps$ with limiting
state space $\mathcal E$ that obeys a LDP with Lagrangian $L$ and
Hamiltonian $H$. Following~\cite{e-ren-vanden-eijnden:2007}, denote by
$\phi(\tau)$, $\tau\in[0,1]$ a \emph{string}, i.e.~a candidate
limiting transition trajectory connecting two fixed points
$a,b\in\mathcal E$. A heteroclinic orbit of the flow $\partial_\theta
H(\phi,0)$ then obeys the relations
\begin{equation}
  \label{eq:ortho}
  \phi(0)=a,\quad\phi(1)=b,\quad \left(\partial_\theta H(\phi(\tau),0)\right)^\perp = 0\,,
\end{equation}
where for a vector field $v(\phi(\tau))$ along the string
$\phi(\tau)$, the notation $v^\perp$ describes the component in the
plane perpendicular to the string,
\begin{align*}
  v(\phi(\tau))^\perp &= v(\phi(\tau)) - \langle v(\phi(\tau)),\gamma(\tau)\rangle \gamma(\tau)\,,\\
  \gamma(\tau)&=|\dot\phi(\tau)|^{-1}\dot\phi(\tau)\,,
\end{align*}
and $\gamma(\tau)$ corresponds to the unit tangent vector along
$\phi$.  Denote by $\phi_n^k$, $n\in\{1,\dots,N\}$, the $k$-th
approximation of the $n$-th image along the discretized string. One
iteration, then consists of two steps.
\begin{enumerate}[(i)]
\item \label{itm:step1} Following proposition~\ref{prop:final},
  integrate the forward gradient flow~(\ref{eq:H-only}),
  $\dot\phi(\tau) = \partial_\theta H(\phi(\tau),0)$ for every image
  along the string via an appropriate integration scheme. For example,
  when choosing forward Euler, set the temporary result
  \begin{equation}
    \label{eq:update}
    \tilde\phi_n = \phi^k_n + \Delta t\, \partial_\theta H(\phi^k_n,0)\,.
  \end{equation}
\end{enumerate}
More sophisticated and higher order time integration schemes are
similarly viable, including implicit ones. In particular, if the
operator with the tightest stability restrictions is linear, it is
generally a good idea to consider exponential time-differencing
schemes~\cite{kassam-trefethen:2005}. In general, all considerations
that are valid for the integration of the limiting dynamics also apply
to the string update step. After applying~(\ref{eq:update}), the
images along the string are no longer distributed in an equidistant
way, and would accumulate at the fixed points $a$ or $b$
if~(\ref{eq:update}) would be repeated indefinitely. Therefore, as a
second step,
\begin{enumerate}[(i)]
  \setcounter{enumi}{1}
\item obtain the next iterate $\phi_n^{k+1}$ by reparametrization of
  $\tilde\phi_n$ by arc-length. The arc-length parameter of the $n$-th
  image is given by the recursive relation
  \begin{equation*}
    s_0 = 0, s_n = s_{n-1} + \|\tilde\phi_n - \tilde\phi_{n-1}\|\,.
  \end{equation*}
  This information can then be used to re-interpolate the images
  $\tilde\phi_n$ to $\phi_n^{k+1}$ at the positions
  \begin{equation*}
    \tilde s_n = n s_N/N\,.
  \end{equation*}
  This ensures that now
  \begin{equation*}
    \|\phi^{k+1}_{n+1} - \phi^{k+1}_{n}\| = \mathrm{cst.}
  \end{equation*}
\end{enumerate}
The re-parametrization step moves points only along the string
$\phi(\tau)$ (up to the accuracy prescribed by the interpolation
scheme). Therefore, after convergence of the algorithm, the remaining
change that occurs in step~(\ref{itm:step1}), is necessarily parallel
to the string, or in other words, at the fixed point of the iteration
the orthogonality condition~(\ref{eq:ortho}) is fulfilled.

\begin{remark}
  All results regarding computational complexity and order of
  convergence of the original `simplified and improved string
  method'~\cite{e-ren-vanden-eijnden:2007} apply also
  here.
\end{remark}

\begin{remark}
  The implementation difficulty of the string method is roughly equal to
  that of the limiting dynamics. Furthermore, we can solve the
  minimization problem with the computational cost of the forward
  integration of the equation, multiplied by the number of
  copies $N$. This is in stark contrast to the full optimization
  problem~(\ref{eq:minimization}) posed by minimizing the original
  rate function, as the corresponding Euler-Lagrange equation is
  second order (in time), the highest order operator of which will be
  the original order squared. For example, if the stochastic PDE in
  question was the stochastic heat equation,
  \begin{equation*}
    \dot\rho =\partial_x^2\rho + \eta,\qquad\EE \left(\eta\eta'\right)=\delta(t-t')\delta(x-x')\,,
  \end{equation*}
  then the corresponding Lagrangian is
  \begin{equation*}
    L(\rho,\dot\rho) = \frac12 \int |\dot\rho - \partial_x^2\rho|^2\,dx
  \end{equation*}
  and the Euler-Lagrange equation, to be solved with suitable boundary
  conditions, is
  \begin{equation*}
    0=\frac{\partial L}{\partial \rho} -\frac {d}{dt} \frac{\partial L}{\partial \dot\rho} = \ddot \rho + \partial_x^4 \rho\,.
  \end{equation*}
  For the string method, one only needs to integrate
  \begin{equation*}
    0 = \dot\rho - \partial_x^2\rho\,.
  \end{equation*}
  Consequently, implementation simplicity and efficiency of the string
  method always exceeds more complex generic optimization methods such
  as the MAM~\cite{e-ren-vanden-eijnden:2004} or
  gMAM~\cite{heymann-vanden-eijnden:2008,
    grafke-schaefer-vanden-eijnden:2017}, and should be preferred
  whenever the reversibility condition holds.
\end{remark}

\section{Examples}
\label{sec:examples}

\subsection{Diffusion in a gradient potential}

As a reminder of the standard setup for which the string method was
originally devised, consider the case of diffusion in a gradient
potential. Let the configuration space $\mathcal E^\eps = \RR^d$,
$\forall\eps>0$, and the system state $X_t\in\mathcal E$ evolve
according to the It\^o SDE
\begin{equation}
  \label{eq:gradient-traditional}
  dX_t^\eps = -\nabla U(X_t^\eps)\,dt + \sqrt{\epsilon}\,dW_t\,.
\end{equation}
Then necessarily $S(x) = U(x) = \frac12 V(x)$, and from
Freidlin-Wentzell theory it follows that
\begin{equation}
  \label{eq:gradient-traditional-L-H}
  L(\psi,\dot\psi) = \frac12|\dot\psi + \nabla U(\psi)|^2\,,\quad
  H(\psi,\theta) = -\langle \theta,\nabla U\rangle + \frac12|\theta|^2\,.
\end{equation}
It is easy to check that indeed the reversibility
condition~(\ref{eq:db-H}) is fulfilled, i.e.~$H(\psi,\nabla U-\theta) =
H(\psi,\nabla U + \theta)$. The minimizers therefore follows
\begin{equation}
  \dot\phi = \pm\nabla U(\phi)\,.
\end{equation}
Numerous applications for this setup exist, most importantly for
chemical systems and molecular dynamics, including Lennard-Jones
clusters, clusters of water molecules, peptides~\cite{wales:2006},
crystal deformations~\cite{e-ren-vanden-eijnden:2005}, and quantum
mechanics/molecular mechanics simulations~\cite{hu-yang:2008}. A
straightforward and well-known generalization of the
flow~(\ref{eq:gradient-traditional}) is to choose an additional
\emph{mobility} matrix $M(x)\in\RR^{d\times d}$ which is symmetric and
divergence free (i.e. $\sum_i \partial_{x_i} M(x) = 0$), so the
evolution equation becomes
\begin{equation}
  \label{eq:SDE-mobility-diffusion}
  dX_t^\eps = -M(X_t^\eps)\nabla U(X_t^\eps)\,dt + \sqrt{\epsilon}M^{1/2}(X_t^\eps)\,dW_t\,,
\end{equation}
where $M^{1/2} (M^{1/2})^T = M$. Then it still holds that $\frac12 V(x) =
U(x) = S(x)$, and Lagrangian and Hamiltonian of
equation~(\ref{eq:gradient-traditional-L-H}) are modified by choosing
the appropriate inner products $\langle u,v\rangle_M = \langle u,
M^{-1} v\rangle$ and $|u|_M = \langle u,u\rangle_M^{1/2}$, with
\begin{align*}
  L(\psi,\dot\psi)&=\frac12|\dot\psi + M(\psi)\nabla U(\psi)|^2_M\,,\\
  H(\psi,\theta) &= -\langle \theta,M(\psi)\nabla U\rangle + \frac12|M(\psi)\theta|^2_M\,.
\end{align*}
In this case, the minimizer obeys the modified relation
\begin{equation}
  \label{eq:mobility-gradient}
  \dot\phi = \pm M(\phi)\nabla U(\phi)\,,
\end{equation}
and the reversibility condition $H(\psi,\nabla U
-\theta)=H(\psi,\nabla U+\theta)$ is
fulfilled. Equations~(\ref{eq:SDE-mobility-diffusion})
and~(\ref{eq:mobility-gradient}) are still within the realm of
classical gradient flows (with mobility). In the next section we will
encounter examples that cannot be phrased even in this form, in
general because the associated Hamiltonian is non-quadratic,
corresponding to a non-Gaussian large deviation principle.

\begin{remark}
  If $M(x)$ is not divergence free, the reversibility condition is
  still applicable if equation~(\ref{eq:SDE-mobility-diffusion})
  additionally is extended by the deterministic drift $\eps\, \mathrm{div}
  M(x)\,dt$, while the LDP remains
  unchanged~\cite{donev-vanden-eijnden:2014}.
\end{remark}

\begin{remark}
  If $M(x)$ is not symmetric, the reversibility condition is
  violated. In this case though, the adjoint dynamics can be written
  down explicitly. Decompose $M = S + A$, with $S=S^T$ and
  $A=-A^T$. Then the adjoint dynamics are
  \begin{equation*}
    dX_t^\eps = -(S-A)\nabla U(X_t^\eps)\,dt + \sqrt{\epsilon}S^{1/2}\,dW_t\,,
  \end{equation*}
  and a modified string method can be built. This is a special case of
  the traverse decomposition~\cite[Chpt 4, Thm
    3.1]{freidlin-wentzell:2012}.
\end{remark}

\subsection{Chemical Reaction Network}
\label{sec:chem-react-netw}

\begin{figure}
  \begin{center}
    \includegraphics[width=196pt]{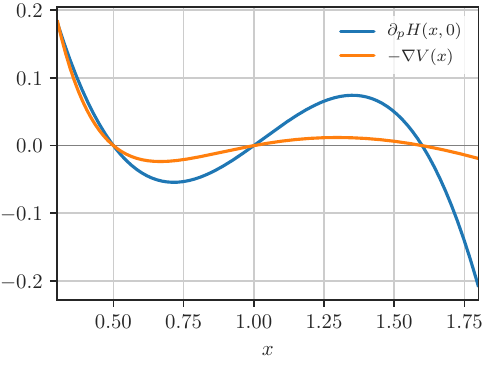}
  \end{center}
  \caption{Comparison of the gradient of the quasipotential, $\nabla
    V$, with the actual generalized gradient drift, $\partial_p H(x,0)
    = \partial_p \psi^\star_x(-\nabla S(x))$ for the chemical reaction
    network of section~\ref{sec:chem-react-netw}. Even though both
    disappear at the two stable fixed points, they disagree in
    magnitude elsewhere. \label{fig:schloegl}}
\end{figure}

In the context of chemical reactions and birth-death processes, one
considers networks of several reactants in a container of volume $D$
which is well-stirred. Take for example the reaction network
\begin{equation*}
  A \stackrel{k_0}{\rightarrow} X\,,\quad
  X \stackrel{k_1}{\rightarrow} A\,,\quad
  2X+B \stackrel{k_2}{\rightarrow} 3X\,,\quad
  3X \stackrel{k_3}{\rightarrow} 2X+B\,,
\end{equation*}
where each process is Poisson with rates $k_i>0$, and where the
concentrations of the reactants $A$ and $B$ are held constant. This
system was introduced in \cite{schloegl:1972}. Even though it can be
considered unrealistic for practical purposes, in particular because
of the presence of ternary reactions, it serves as a toy model to
understand bi-stable reaction networks.  Its dynamics can be modeled
as a MJP on $\ZZ_+$ with generator
\begin{equation}
  \label{eq:prob22}
  (\mathcal L f)(n) = A(n) \left(f(n+1) - f(n)\right) + B(n)\left(f(n-1) - f(n)\right)
\end{equation}
and with the propensity functions
\begin{align*}
  A(n) &= k_0 D + (k_2/D) n(n-1)\\
  B(n) &= k_1 n + (k_3/D^2) n(n-1)(n-2)\,.
\end{align*}
The model above satisfies a large deviation principle in the following
scaling limit: Denote by $c=n/D$ the concentration of $X$, and
normalize it by a typical concentration $c_0$, so that $x=c/c_0$. Now,
for a large number of particles per lattice site $c_0
D=\epsilon^{-1}$, we obtain to leading order
\begin{equation}
  \label{eq:prob22scaled}
  \begin{aligned}
    (\mathcal L^\epsilon f)(x) &= \epsilon^{-1} \Big(a(x)
    \left(f(x+\epsilon)
    -f(x)\right) \\&\qquad\quad+
    b(x)\left(f(x-\epsilon)-f(x)\right)\Big)\,,
  \end{aligned}
\end{equation}
on $\mathcal E^\eps = \eps \ZZ_+$, where we defined
$k_i=\lambda_i(c_0)^{1-i}$, and
\begin{equation*}
  a(x) = \lambda_0 + \lambda_2 x^2\,,\quad
  b(x) = \lambda_1 x + \lambda_3 x^3\,.
\end{equation*}
The large deviation principle for~(\ref{eq:prob22scaled}) can be
formally obtained via WKB analysis which gives a Hamilton-Jacobi
operator associated with a Hamiltonian that is also the one rigorously
derived in LDT~\cite{shwartz-weiss:1995},
\begin{equation*}
  H(x,p) = a(x) \left(e^p -1\right) + b(x) \left(e^{-p} -1\right)\,.
\end{equation*}
Note that this Hamiltonian is not quadratic in its conjugate momentum,
implying that no Gaussian SDE exists (including no diffusion in a
potential) with the same large deviation principle. Detailed balance
is nevertheless fulfilled, by realizing that
\begin{equation}
  \label{eq:schloegl-gradV}
  \nabla V(x) = \log \frac {b(x)}{a(x)}\,,
\end{equation}
under which the reversibility condition $H(x,p) = H(x,\nabla V-p)$
is confirmed.  The large deviation minimizers can therefore explicitly
be computed to be
\begin{equation*}
  \dot x = \pm \partial_p H(x,0) = \pm (a(x) - b(x))\,.
\end{equation*}
This precisely corresponds to the law of mass-action of the reaction
network, as well as its time-reversed variant. 

The generalized gradient flow for this system is explicitly given by
\begin{equation*}
  \dot x = \partial_p \psi^\star_x(-\nabla S)\,,
\end{equation*}
with $\nabla S = \frac12 \nabla V$ given by~(\ref{eq:schloegl-gradV}) and
\begin{equation*}
  \psi^\star_x(p) = \sqrt{ab} \left(e^p-1\right) + \sqrt{ab}\left(e^{-p}-1\right)\,.
\end{equation*}
Figure~\ref{fig:schloegl} highlights the fact that these dynamics are
very different from the gradient of the potential itself: Even though
both disappear at the fixed points, their behavior away from the fixed
points does not agree.

\subsection{Zero-range process}

\begin{figure}
  \begin{center}
    \includegraphics[width=234pt]{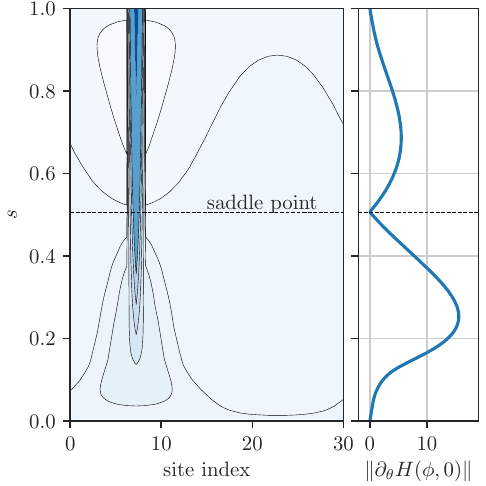}
  \end{center}
  \caption{String of the zero-range process towards condensation in a
    single lattice site $i=8$ (left): For each $s\in[0,1]$ along the
    string, the configuration $\vec\rho(s)$ is depicted, starting from
    a constant density at $s=0$ and reaching a condensate at $i=8$ for
    $s=1$. The corresponding strength of the generalized gradient flow
    (right), identifying the critical nucleus at the saddle, where
    $\|\partial_\theta H(\phi,0)\|=0$. \label{fig:zrp-string}}
\end{figure}

Consider a lattice gas on a one dimensional lattice with $L\in\NN$
sites, where on each lattice site $i$ there are $n_i$ particles. The
system state is described by $\vec n\in \ZZ_+^L$. Particles can hop to
neighboring sites on the left or right with a rate $\gamma(n_i)$
depending only on the local occupation number, so that the total
number of particles $N$ is conserved, $N=\sum_{i=1}^L n_i$. Such a
system is called a zero-range process (ZRP), which is a MJP with
generator
\begin{equation*}
  \mathcal L f(\vec n) = \sum_{i=1}^L \gamma(n_i) \left(f(\vec n + \vec e_i^+) + f(\vec n + \vec e_i^-) - 2f(\vec n)\right)\,,
\end{equation*}
where $\vec e_i^\pm$ is the vector with zero entries everywhere,
except $-1$ at $i$ and $1$ at $i\pm1$. For finite $L$, a large
deviation principle can be obtained for this MJP in $\eps=N^{-1}\to0$
by considering the rescaled quantity $\rho_i = \bar\rho n_i/N = \eps
n_i$, $\bar\rho>0$, for $\mathcal E^\eps=\eps\ZZ_+^L$ and $\mathcal E
= \RR_+^L$. After furthermore rescaling time, the generator then reads
\begin{equation*}
  \mathcal L^\eps f(\vec \rho) = \eps^{-1}\sum_{i=1}^L \gamma(\rho_i) \left(f(\vec \rho + \eps\vec e_i^+) + f(\vec \rho + \eps\vec e_i^-) - 2f(\vec \rho)\right)\,,
\end{equation*}
where the jump rates $\gamma$ have been rescaled appropriately
\cite{feng-kurtz:2006, bertini-de_sole-gabrielli-etal:2015}. The large
deviation Hamiltonian reads
\begin{equation}
  \label{eq:H-zrp}
  H(\rho,\theta) = \sum_i \gamma(\rho_i) \left(e^{\theta_{i-1}-\theta_i} + e^{\theta_{i+1}-\theta_i}
  - 2\right)\,.
\end{equation}
Note again that the Hamiltonian is non-quadratic in the conjugate
momentum, meaning that no Gaussian SDE can be found with large
deviation Hamiltonian~(\ref{eq:H-zrp}). The system is not in detailed
balance in general, but one can choose the rates $\gamma(x)$ in order
to enforce reversibility. The reversibility condition amounts to
\begin{equation}
  \label{eq:5}
  \frac{P(\vec\rho\to \vec\rho+\eps \vec e_i^+)}{P(\vec\rho+\eps \vec e_i^+\to \vec\rho)} = \frac{\rho_\infty(\vec\rho+\eps \vec e_i^+)}{\rho_\infty(\vec\rho)}
\end{equation}
i.e.~the ratio of forward and backward reaction rates has to
correspond to the relative probability of the respective states. Since
the density $\rho_\infty$ is connected to the quasipotential via
$\rho_\infty(\vec\rho) \asymp \exp(-\eps^{-1} V(\vec\rho))$ as per
section~\ref{sec:large-devi-princ}, and furthermore $P(\vec\rho\to
\vec\rho+\eps \vec e_i^+)=\gamma(\rho_i)$, and $P(\vec\rho+\eps \vec
e_i^+ \to \vec\rho)=\gamma(\rho_{i+1})$, the reversibility
condition~(\ref{eq:5}) for $\eps\to0$ translates to
\begin{equation*}
  \frac{\gamma(\rho_{i+1})}{\gamma(\rho_i)} = \exp(-(\nabla_i V(\vec\rho)-\nabla_{i+1} V(\vec\rho)))\,.
\end{equation*}
A possible choice to fulfill this constraint is
\begin{equation}
  \label{eq:zrp-gradV}
  \nabla_i V(\vec\rho) = \ln \gamma(\rho_i) + C\,,
\end{equation}
where the constant $C$ is fixed by the conserved mean density
$\bar\rho=L^{-1}\sum_{i}\rho_i$ via $C=-\ln \gamma(\bar\rho)$. We
therefore obtain
\begin{equation*}
  V(\vec\rho) = \sum_i \int_0^{\rho_i} (\ln\gamma(y)+C)\,dy\,,
\end{equation*}
which is the correct potential for the
ZRP~\cite{grosskinsky-schutz-spohn:2003}. In particular, for
non-interacting particles, we have $\gamma(x)=x$ and
\begin{equation*}
  V(\vec\rho) = \sum_i (\rho_i\ln\left(\frac{\rho_i}{\bar\rho}\right) - \rho_i)\,.
\end{equation*}
For general $\gamma(x)$ obeying the reversibility condition, the
transition trajectories now follow
\begin{equation}
  \label{eq:ZRP-gradient}
  \dot \rho_i = \pm \partial_{\theta_i} H(\vec\rho,0) = \pm(\gamma(\rho_{i-1}) + \gamma(\rho_{i+1}) -2\gamma(\rho_{i}))\,.
\end{equation}
To highlight that this flow again is a generalized gradient flow, note
that (\ref{eq:ZRP-gradient}) is of the form
\begin{equation*}
  \dot \rho_i = \partial_{\theta_i} \psi^\star_\rho(-\nabla S(\vec\rho))\,,
\end{equation*}
with $\nabla S(\rho) = \frac12 \nabla V(\rho)$ given by~(\ref{eq:zrp-gradV}) and
\begin{align*}
  \psi^\star_\rho(\vec\theta) &= \sum_i \sqrt{\gamma_{i}\gamma_{i-1}}\left(e^{\theta_{i-1}-\theta_i}-1\right)\\ &+ \sum_i \sqrt{\gamma_i\gamma_{i+1}}\left(e^{\theta_{i+1}-\theta_i}-1\right)\,,
\end{align*}
where $\gamma_i = \gamma(\rho_i)$.

Notably, for specific choices of $\gamma(x)$, this system has multiple
stable fixed points. For example, taking $\gamma(x)=D+\exp(-x)$, for
$D=\frac1{10}$ and $\bar\rho=1$ provides two stable fixed points, one
being a constant solution at $\rho_i=\bar\rho$ and one being a
\emph{condensate}, where a macroscopic fraction of the density is
concentrated in a single lattice site. This setup is metastable in that
fluctuations from the finiteness of the number of particles will
eventually force the system from one fixed-point to the other, the
transition happening along the heteroclinic orbits of
equation~(\ref{eq:ZRP-gradient}) in the limit $N\to\infty$. The
corresponding transition trajectory is depicted in
figure~\ref{fig:zrp-string} (left). Here, the $x$-axis represents the
site index, in this particular case from $0$ to $30$ with $L=31$. The
$y$-axis represents the parameter $s$ along the string. The shading of
the contour plot depicts each configuration $\vec\rho(s)$ along the
string, starting at the constant density at $s=0$, and transitioning
into the condensate at $s=1$. The relevant saddle is reached at
$s=0.5$, marked as a dashed line in the contour plot, and identified
through the point at which $\|\partial_\theta H(\phi,0)\|=0$, depicted
in figure~\ref{fig:zrp-string} (right). Notably, a spatially extended
condensate forms into a \emph{critical nucleus}, which is reached at
the saddle point.

\subsection{Interacting particle system}

\begin{figure}
  \begin{center}
    \includegraphics[width=234pt]{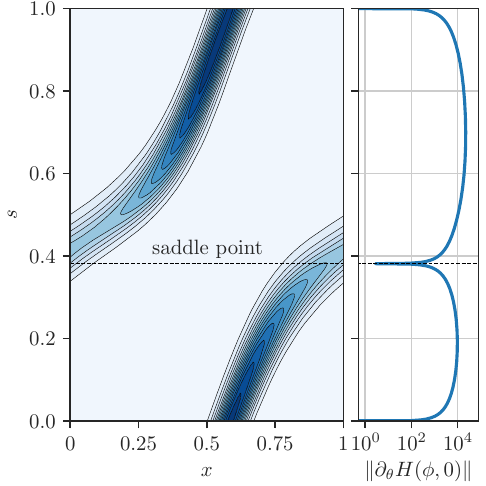}
  \end{center}
  \caption{String of the interacting particle system, where the
    cluster of attractive particles performs one revolution around the
    periodic domain (left). For each $s\in[0,1]$, the configuration
    $\rho(s,x)$ is depicted. The corresponding strength of the
    Wasserstein gradient flow identifies the relevant saddle point
    (right). \label{fig:particles-string}}
\end{figure}

On a periodic domain $\Omega=[0,1]$, consider a system of $N$
interacting Brownian particles at $X_i\in\Omega$, $i\in\{1,\dots,N\}$,
in a potential $U(x) : \Omega \to \RR$ and with interaction potential
$K(x)$. Each particle is modeled by its own It\^o diffusion,
\begin{align*}
  dX_i(t) =& -\nabla U(X_i(t))\,dt\\& - \frac1N \sum_{j=1}^N \nabla K(X_i(t)-X_j(t))\,dt + \sqrt{2}\,dW_i(t)\,.
\end{align*}
The hydrodynamic limit of this system is
\begin{equation*}
  \partial_t \rho = \partial_x^2 \rho + \partial_x\left(\rho \partial_x U + \partial_x(\rho\star K)\right)\,.
\end{equation*}
As discussed in~\cite{adams-dirr-peletier-etal:2013}, this can be
interpreted as a generalized gradient flow in the ($\rho$-dependent) Wasserstein metric, evolving according to
\begin{equation*}
\partial_t \rho = -M(\rho)\nabla_\rho V, \quad M(\rho)\xi = -\nabla\cdot(\rho\nabla \xi)
\end{equation*}
with
\begin{equation*}
  V(\rho) = \int\left(\rho\log\rho - \rho + \rho U + \frac12\rho (K\star\rho)\right)\,dx\,.
\end{equation*}
On the other hand, from a large deviation perspective, the Hamiltonian
of this system reads
\begin{align*}
  H(\rho,\theta) &= \int_0^1\!\!\left( \theta\partial_x^2\rho + \theta \partial_x(\rho\partial_x U + \partial_x (\rho\star K)) - \rho (\partial_x \theta)^2\right)dx\\
  &= \langle -M(\rho) \nabla_\rho V(\rho), \theta\rangle +  \langle \theta, M(\rho) \theta\rangle\,,
\end{align*}
where the inner product and norm are $L^2$. As concrete demonstration
of our algorithm, we take a periodic potential $U(x) = \alpha
\cos(2\pi x)$, which has a unique minimum at the center of the
domain. As interaction potential, we pick $K(x)$ such that $\partial_x
K(x) = w(x-\delta)$, with
\begin{equation*}
  w(x) =
  \left\{ \begin{array}{@{\kern2.5pt}lL}
    \hfill\beta x \exp\left(-\frac{2x^2}{1-2x^2}\right) & $x\le\frac12$\\
    \hfill 0&else,
  \end{array}\right.
\end{equation*}
which results in a locally parabolic interaction which tapers off to
$0$. Notably, for $\delta\ne0$, the interaction potential is not
symmetric, resulting in an effective net force, e.g.~to the right for
$\delta>0$. In total, particles tend to stick together and try to move
right, but collect within the basin of the potential. As a
consequence, the system is not meta-stable, and the unique fixed point
is a cluster of particles slightly off $x=\frac12$. We can
nevertheless compute an interesting transition trajectory: We can
force the particle cloud to revolve once around the periodic domain,
i.e.~ask for the most likely trajectory that leads to the particles
collectively traveling up the barrier of $U(x)$ towards $x=1$, and
down again from $x=0$ towards the fixed point. The resulting string is
depicted in figure~\ref{fig:particles-string} (left). Again, the
contour plot represents the configurations $\rho(x)$ along the string
parameter $s\in[0,1]$, starting and ending at the same (fixed point)
configuration, but wrapping around the periodic domain once. The
corresponding strength of the gradient drift is depicted in
figure~\ref{fig:particles-string} (right), which identifies the
relevant saddle for the transition. Concretely, we choose
$\alpha=0.5\cdot 10^3$, $\beta=0.5\cdot 10^2$, and $\delta=5\cdot
10^{-2}$.

\section{Concluding remarks}

In this paper, we showed how the fact that a large deviation principle
induces a generalized gradient flow for reversible processes can be
used to obtain geometric properties of limiting transition
trajectories between fixed points. In particular, we showed that under
suitable conditions every minimizer of a large deviation principle of
a reversible process can be interpreted as a heteroclinic orbit (or
its time-reverse) in a generalized gradient flow.

This fact has important consequences for the numerical computation of
the most likely transition trajectories. In particular, the string
method, originally devised to effectively compute minimizing
trajectories for diffusions in a potential landscape, can be adapted
to the wider class of generalized gradient flows.

We demonstrated the feasibility of this approach by computing
transition trajectories for the condensation of a zero range process, a
particular lattice gas model, as well as the hydrodynamic limit of
interacting particles, in all cases computing the most likely
trajectory realizing a certain event, and identifying the saddle point
along the transition, at which the dynamics cross between the
different basins of attraction.

\section*{Acknowledgments}

TG would like to thank M.~Peletier, A.~Montefusco, E.~Vanden-Eijnden,
and H.~Touchette for fruitful discussions.


\bibliographystyle{iopart-num}
\bibliography{bib}

\end{document}